%% file: main.tex
\documentclass{article}
\usepackage{caption}
\usepackage{graphicx}
\usepackage{pgfplots}
\usepackage{amsmath}
\usepackage[hyphens]{url}
\pgfplotsset{compat=1.18}
\usepackage{booktabs}
\usepackage{xcolor}
\usepackage[
backend=biber,
style=nature,
sorting=none,
defernumbers=true
]{biblatex}
\usepackage{xcolor} 
\input{utils} 
\usepackage{authblk}

\usepackage[hidelinks,colorlinks=true]{hyperref}
\usepackage[capitalize,noabbrev]{cleveref}
\usepackage{multirow}

\title{Benchmarking \textbf{\simulacra}'s Quantum Accurate Synthetic Data Generation for Chemical Sciences}

\author[1]{Fabio Falcioni\thanks{Corresponding author(s): \href{mailto:fabio.falcioni@simulacra-ai.com}{fabio.falcioni@simulacra-ai.com}, \href{mailto:aleksei@simulacra-ai.com}{aleksei@simulacra-ai.com}, \newline \indent \textcopyright~2025 Simulacra Research Inc.\ All rights reserved.}}
\author[1]{Elena Orlova}
\author[1,2]{Timothy Heightman}
\author[1]{Philip Mantrov}
\author[1]{Aleksei Ustimenko$^*$}

\affil[1]{Simulacra Research Inc., London, UK \& Chicago, USA}
\affil[2]{ICFO-Institut  de  Ciències  Fotòniques,  The  Barcelona  Institute  of  Science  and  Technology, 08860 Castelldefels (Barcelona), Spain}

\date{\today}

\begin{document}
\maketitle

\begin{abstract}
In this work, we benchmark \simulacra's synthetic data generation pipeline against a state-of-the-art Microsoft pipeline on a dataset of small to large systems. By analyzing the energy quality, autocorrelation times, and effective sample size, our findings show that \simulacra's Large Wavefunction Models (LWM) pipeline, paired with state-of-the-art Variational Monte Carlo (VMC) sampling algorithms, reduces data generation costs by 15-50x, while maintaining parity in energy accuracy, and 2-3x compared to traditional CCSD methods on the scale of amino acids. This enables the creation of affordable, large-scale \textit{ab-initio} datasets, accelerating AI-driven optimization and discovery in the pharmaceutical industry and beyond. Our improvements are based on a novel and proprietary sampling scheme called Replica Exchange with Langevin Adaptive eXploration (RELAX).
\end{abstract}
\section{Introduction}
Many industries have recently adopted different types of Artificial Intelligence (AI) models, more specifically Large Language Models (LLMs)\cite{raza2025industrial}. However, it is well known that LLMs can be unreliable \cite{zhou2024larger} as they suffer from various phenomena such as hallucinations, catastrophic forgetting, and model collapse, to name a few \cite{myers2024foundation}. As artificial intelligence advances into high-stakes areas such as drug discovery, reliability shifts from a performance metric to a non-negotiable requirement. Unlike text or image synthesis, errors in molecular prediction propagate into failed experiments, wasted capital, and potential harm to patients. It is here that technology and science, grounded in physics, mathematics, and chemistry, have been employed for many decades. For example, computational chemistry and bioinformatics have been staples of the pharmaceutical industry, where large-scale virtual screening, high-throughput screening, and molecular modeling are used to select viable drug candidates for all kinds of rare and common diseases.

However, AI has also reshaped core tasks in these sectors over the last five years from protein structure prediction and de-novo protein design to structure-based drug discovery and quantum-accurate molecular simulations. Protein folding systems such as AlphaFold, RoseTTAFold, and ESMFold have expanded structural coverage to unprecedented scales, improving downstream hypothesis generation in structural biology and target assessment \cite{jumper2021alphafold,varadi2022alphafold,baek2021rosetta,lin2023esmfold}. Generative and equivariant models now propose binders and macromolecular assemblies (e.g., RFdiffusion\cite{watson2023RFDiffusion}) and frame docking as a diffusion process over poses (e.g., DiffDock\cite{corso2022diffdock}), accelerating early-stage design. At the atomistic level, physically informed machine-learning interatomic potentials (MLIP) embed symmetries and Quantum Mechanics (QM) to deliver \textit{ab-initio}-like accuracy at classical simulation speeds and length scales that were previously beyond reach \cite{batatia2022mace, batzner2022nquip, smith2017ani, batatia2023foundation}.

However, despite these advances, data quality remains the dominant bottleneck in terms of reliability and physicality. Most openly available datasets that power molecular modeling at scale are computed using approximations to solve the Schr\"odinger's equation, such as DFT. This includes many of the most impactful recent releases: SPICE for drug-like chemistry and biomolecular fragments\cite{eastman2023spice}, QMugs for pharmacologically relevant small molecules\cite{isert2022qmugs}, and Meta FAIR’s Open Molecules 2025 (OMol25)\cite{levine2025omol}, a massive dataset reported to comprise more than 100 million DFT calculations at the $\omega$B97M-V/def2-TZVPD level, spanning 83 elements, diverse charge/spin states, explicit solvation, and reactive structures. The scale and breadth of these resources are a leap forward, but they are still based on the inaccurate DFT method, and therefore inherit both systematic and unsystematic errors of the chosen level of theory.
These methods are nowadays employed in industry to obtain molecular properties and train other downstream machine learning models. However, they are not uniformly accurate across chemistry. Even sophisticated DFT functionals (e.g.,  $\omega$B97M-V with non-local correlation) can struggle in regimes central to drug discovery and catalysis. For example, in long-range charge transfer, delicate non-covalent interactions, open-shell and multi-reference transition-metal complexes, spin-splitting energetics, and strongly correlated bonding. Benchmark studies (GMTKN55; extensive functional surveys \cite{goerigk2017gmtk55}) and focused reviews repeatedly document these limitations and their variability across chemical space; models trained exclusively on DFT inherit these biases \cite{mardirossian2017dftreview, kuryla2025benchDFTforces}.
In contrast, wavefunction (``post-H'') methods such as Coupled Cluster Single-Doubles-(Triple) (CCSD(T)) or CBS (complete-basis-set limit), and multi-reference treatments when required, more faithfully solve the electronic Schr\"odinger equation and set the standard for predictive thermochemistry, kinetics, and intermolecular interactions.

It turns out, however, that even the availability of such high-accuracy methods, which provide nearly exact results, does not solve the underlying issue of scalability for large amounts of data in terms of quantity and system size. The so-called ``gold-standard'' methods of quantum chemistry scale as $\mathcal{O}(N^7)$ (where $N$ is the number of one-electron basis functions that represent the system), making them prohibitively expensive \cite{doi:10.1021/acs.jctc.4c01314}. For molecules with up to 32 atoms, generating $10^5$ data points (e.g., conformations) using CCSD(T) on commonly used platforms such as ORCA \cite{https://doi.org/10.1002/wcms.1606} or Q-Chem \cite{10.1063/5.0055522} can cost millions of dollars in compute resources. For larger systems ($> 32$ atoms), such as peptides, small proteins, drug complexes or materials' unit cells, the cost becomes astronomical, limiting the most recent datasets to low-fidelity data.
Therefore, while OMol25 and Meta’s broader open programs (e.g., the Open Catalyst datasets built with DFT for surfaces and adsorbates \cite{levine2025omol, chanussot2021ocp, sahoo2025OC25}) are invaluable stepping stones for training foundation models, their trustworthiness for high-stakes predictions is ultimately limited by the underlying method employed to obtain the ``ground truth".

To close this gap, a complementary path, made possible by the latest advancements in AI technology has emerged, hereby referred to as Large Wavefunction Models (LWMs). LWMs are foundation neural-network wavefunctions optimized by Variational Monte Carlo (VMC) that directly approximate the many-electron wavefunction. Unlike DFT and post-HF methods, these models are trained by minimising the variational energy, yielding upper bounds that approach the exact Born–Oppenheimer solution as the \singleansatz becomes more expressive, and providing unbiased estimators for observables (densities, energies, forces, dipoles).
In practice, they capture both static and dynamic correlation without hand-crafted functionals. Recent results extend beyond small test cases. For example, neural wavefunctions achieve state-of-the-art ground-state energies, tackle excited states with new VMC principles and handle positronic complexes and strongly-correlated superfluids\cite{pfau2020ferminet,pfau2024ferminetexcitedstates,cassella2024positronic}. They are also becoming transferable across multiple molecules and even solids, pointing to the feasibility of pretrained ``foundation'' wavefunctions that fine-tune rapidly to new systems\cite{scherbela2024taowfn, gao2021pesnet1, gerard2024transferablewfnsolids, foster2025abinitiofoundationmodel}. In effect, this route provides physically trustworthy data labels by (variationally) solving the Schr\"odinger equation, addressing the data-quality limitations that constrain DFT-only datasets like OMol25, and offering a path to seed $\Delta$-learning targets \cite{ramakrishnan2015big}, calibrate MLIP potentials, ML-DFT functionals, and establish gold-standard benchmarks for regimes where DFT or even post-HF methods are unreliable (e.g., charge transfer, non-covalent interactions, open-shell/transition-metal chemistry, to name a few).

The overall effect of using such physics-driven models puts drug and materials industries on a firmer physical footing by supplying gold-standard labels at a fraction of the cost. In pharma, that means scoring functions and force fields for molecular modelling and simulations that include polarization and charge-transfer correctly, improving pose ranking, covalent warhead barrier predictions, and excited-state design (fluorophores, photo-switches) where current methods often fail. In materials, it sharpens adsorption energies and rate-limiting barriers for (electro)catalysis, fixes polaron/redox energetics in battery electrodes, and yields more reliable singlet–triplet gaps for emitters. Practically, these data enable $\Delta$-learning to lift the currently deployed AI models in the regimes that matter, reducing the need for ever-larger but noisy datasets. The result is faster, more trustworthy downstream selection, and fewer costly surprises in the lab.

In this work, we present an advance in \simulacra technology that unlocks the potential of Large Wavefunction Models (LWMs) by accelerating a critical subroutine applicable to training, finetuning and evaluation: sampling schemes. Our main result is a systematic decrease in autocorrelation time that cuts costs over our test sets by an average of 28x, ranging from 15-50x, when comparing to \microsoft pipeline \cite{foster2025abinitiofoundationmodel}. We note that there is systematic reluctance in the literature of accounting for autocorrelation time correction in analyses of VMC. This is likely because up until now, most research has assumed computation of one system at a time, rather than employing LWMs as generalized wavefunctions to be pretrained and fine-tuned for query-based research of chemical properties.
We begin with essential background on quantum chemistry fundamentals and the well-established Variational Monte Carlo (VMC) framework, then introduce LWMs as a generalization of neural-network wavefunctions (NN-WFNs). We next describe the RELAX sampling algorithm and show how it enables scalable LWMs. Finally, we benchmark our framework for efficiency and cost against \microsoft’s state-of-the-art, using the same LWM architecture (OrbFormer \cite{foster2025abinitiofoundationmodel} pretrained on the Light-Atom-Curriculum dataset) in both pipelines to ensure a like-for-like comparison.

\section{Background: Quantum Chemistry Fundamentals}
The core of quantum chemistry is the resolution of the famous time-independent Schr\"odinger equation
\begin{equation}
    \hat{H} \Psi = E \Psi 
\end{equation}
where $\hat{H}$ is the Hamiltonian operator of a many-body interacting system of atoms and electrons, $E$ is the total ground state energy of the system, and $\Psi$ is the wavefunction.
This is what enables the prediction of useful molecular properties. In this section, we briefly introduce some concepts relevant to quantum chemistry and methodologies employed in industry, along with their theoretical and practical pitfalls. We begin with the Born-Oppenheimer approximation, followed by a brief overview of Density Functional Theory (DFT), Coupled Cluster (CC), and Variational Monte Carlo (VMC) methods.

\subsection{Born-Oppenheimer Approximation}
The Born-Oppenheimer approximation separates nuclear and electronic degrees of freedom due to the mass difference between nuclei and electrons \cite{https://doi.org/10.1002/andp.19273892002}. In this approximation, a molecular Hamiltonian reads
\begin{equation}
\hat{H} = \hat{T}_n + \hat{T}_e + \hat{V}_{nn} + \hat{V}_{ee} + \hat{V}_{ne},
\end{equation}
where $\hat{T}_n, \hat{T}_e$ are the kinetic energies of nuclei and electrons, and $\hat{V}_{nn}, \hat{V}_{ee}, \hat{V}_{ne}$ are the potential interactions between nuclei and electrons. Fixing the nuclei positions $\mathbf{R}$, the electronic Hamiltonian becomes:
\begin{equation}
\hat{H}_{el} = \hat{T}_e + \hat{V}_{ee} + \hat{V}_{ne}(\mathbf{R}) + \hat{V}_{nn}(\mathbf{R}).
\label{eq:elec_ham}
\end{equation}
Solving the time-independent Schr\"odinger equation for a set of nuclear coordinates:
\begin{equation}
\hat{H}_{el} \Psi_{el}(\mathbf{r}; \mathbf{R}) = E_{el}(\mathbf{R}) \Psi_{el}(\mathbf{r}; \mathbf{R}),
\label{eq:schrodinger2}
\end{equation}
yields the total electronic energy of a system $E_{el}(\mathbf{R})$.
Solving \Cref{eq:schrodinger2} for multiple sets of coordinates defines the Potential Energy Surface (PES) of a system, which can be used to explain chemical reactivity and nuclear dynamics in simulations.
It is under the Born-Oppenheimer approximation that most of today's quantum chemistry is performed. This has led to the development of methods such as Hartree-Fock (HF), DFT, and post-HF to compute approximate values of $E_{el}(\mathbf{R})$ under various further approximations, whose theory we briefly review below. 

\subsection{Density Functional Theory and Coupled-Cluster}
The Hohenberg-Kohn theory states that there exists a universal functional $E[\rho]$ of the ground-state electron density $\rho(\mathbf{r})$ that yields the exact ground-state energy at its minimum \cite{hohenberg1964inhomogeneous,parr1989density}. Kohn-Sham (KS) DFT \cite{kohn1965KSDFT} makes this practical by introducing a fictitious non-interacting system with single-particle orbitals ${\phi}_i(\mathbf{r})$ whose density reproduces the interacting one $\rho(\mathbf{r}) = \sum_i f_i |\phi_i(\mathbf{r})|^2$, with occupations $f_i$. The KS energy is therefore obtained as 
\begin{equation}
    E[\rho] = T_s[\rho] + E_H[\rho] + E_{xc}[\rho] + \int v_{ext}(\mathbf{r})\rho(\mathbf{r})d\mathbf{r},
\end{equation}
where $T_s[\rho]$ is the kinetic energy of the non-interacting KS system, $E_H[\rho] = \frac{1}{2}\int\int \frac{\rho(\mathbf{r})\rho(\mathbf{r'})}{|\mathbf{r} - \mathbf{r}'|}$ is the Hartree energy (classical Coulomb), $E_{xc}[\rho]$ is the exchange-correlation energy that gathers all the many-body effects missing from $T_s$ and $E_H$, and finally $v_{ext}(\mathbf{r}) = -\sum_A Z_A/|\mathbf{r} - \mathbf{R}_A|$ is  the electron-nuclei interaction, which acts as an external potential on the interacting system.
Functional minimization leads to the KS one-electron equations, similar to Schr\"odinger's equation:
\begin{equation}
    \left [ \frac{1}{2}\nabla^2 + v_{ext}(\mathbf{r}) + v_H[\rho](\mathbf{r}) + v_{xc}[\rho](\mathbf{r})\right]\phi_i(\mathbf{r}) = \epsilon_i\phi_i(\mathbf{r}),
\end{equation}
with $v_H[\rho](\mathbf{r}) = \delta E_H / \delta \rho$ and $v_{xc}[\rho](\mathbf{r}) = \delta E_{xc} / \delta \rho$, which is referred to as an exchange-correlation (XC) potential. 

In practice, one chooses an approximate exchange-correlation functional $E_{xc}$ (e.g., LDA, GGA, meta-GGA, hybrid, double-hybrid, see \cite{parr1989DFTReview} for further detail), expands $\{\phi_i\}$ in a basis (e.g., Gaussian atomic orbitals), and iterates to self-consistency. The output of such calculations includes the electron density, total energy and, with further computation, forces and vibrational frequency, to name a few \cite{parr1989DFTReview}. However, DFT comes with limitations, the biggest of these being that the exact XC functional $E_{xc}$ is unknown.

Approximations for the XC potential can be devised; however, they inherit self-interaction, delocalization errors, and lack of derivative discontinuity \cite{cohen2008DFTLimitations, mardirossian2017dftreview}. This can result in band gaps, charge-transfer, and reaction barrier energies that are orders of magnitude away from the actual values \cite{kieron2013DFTErrors1, Cohen2012DFTChallenges}.
The practical details of DFT implementations also matter. For example, the choice of numerical quadrature grids is not straightforward and can significantly alter results \cite{murray1993dftgrids}. Dispersion effects need to be included empirically (D3/D4 dispersion corrections\cite{grimme2011D3}) or via non-local kernels (VV10/rVV10\cite{sabatini2013VV10}). Basis-set incompleteness and Basis-Set Superposition Error (BSSE) also largely affect results \cite{mentel2014BSSE}. 
DFT granularity and sensitivity to context make it ill-suited for massive, diverse datasets, as systematic plausibility validation across such breadth is infeasible.

Next, we turn our attention to the Coupled Cluster (CC) method \cite{bartlett2007CoupledCluster, shavitt2009ManyBodyMethods}, the most well-known method in the family of wavefunction methods that directly target the many-electron state $|\Psi \rangle$. CC starts from a single-determinant reference (usually the Hartree-Fock wavefunction) $|\Phi_0 \rangle$ and the exponential \singleansatz:
\begin{equation}
    |\Psi_{CC}\rangle = e^T|\Phi_0\rangle, \qquad T = T_1 + T_2 + T_3 + ...,
\end{equation}
where $T_k$ is an operator that promotes $k$ electrons from occupied to virtual spin-orbitals, with amplitudes $\{t_\mu\}$. The similarity-transformed Hamiltonian $\bar{H} = e^{-T}\hat{H}e^T$ yields the energy and amplitude equations:
\begin{equation}
    E_{CC} = \langle \Phi_0 | \bar{H}|\Phi_0\rangle, \qquad \langle \Phi_\mu | \bar{H}|\Phi_0\rangle = 0  \quad (\mu \neq 0).
\end{equation}
Truncating CC at single and double excitations gives CCSD, while adding a non-iterative perturbative triples correction gives CCSD(T). Due to the exponential form, the CC is \textit{size-extensive} and rapidly convergent for weak to moderate correlation. Near equilibrium, for single-reference closed-shell molecules, CCSD(T) often delivers excellent thermochemistry and barrier heights, hence its reputation as the ``gold-standard" of quantum chemistry methods \cite{Rezac2013CCSDTGoldStandard, bartlett2007CoupledCluster}.

However, CCSD scales as $\mathcal{O}(N_{basis}^6)$ and CCSD(T) as $\mathcal{O}(N_{basis}^7)$, with $N_{basis}$ number of one-particle basis functions. As such, it has a large memory footprint and reaching a complete-basis set limit (CBS) remains infeasibly expensive. More fundamentally, CCSD(T) is \textit{non-variational} (i.e. energies are not guaranteed upper bounds), and it is not reliable when static correlation is present. The latter is common in many phenomena, such as bond dissociation, diradicals/polyradicals chemistry, transition-metal and f-element chemistry, because the single-determinant reference  $|\Phi_0 \rangle$ is qualitatively wrong.
Thus, CCSD(T) is a strong conditional standard (single-reference, near equilibrium), but not a universal one \cite{crawford2019ReducedCoupledCluster, margraf2017CCReview}. This makes the case for the use of more sophisticated methods such as VMC, discussed briefly below.

\subsection{Variational Monte Carlo}
\label{sec:VMC}

Instead of integrating deterministically as in previously described methods, Monte Carlo methods estimate the properties of a many-body electronic system by sampling electronic configurations $\mathbf{r} = (\mathbf{r}_1, \ldots, \mathbf{r}_{N_c}$) drawn from the sampling distribution
\(\pi(\mathbf r)\propto |\Psi(\mathbf r)|^2\). The variational component of Variational Monte Carlo (VMC) comes from variationally tuning the parameters of a trial wavefunction $\Psi$, used to generate samples, in order to produce better estimates. More concretely, we can define the objective functional
\begin{equation}
E[\Psi]
= \frac{\langle \Psi | \hat H_{el} | \Psi \rangle}{\langle \Psi | \Psi \rangle}
= \frac{\int \Psi^*(\mathbf r)\,\hat H_{el}\Psi(\mathbf r)\,d\mathbf r}{\int |\Psi(\mathbf r)|^2\,d\mathbf r}
= \int E_{L}(\mathbf{r})\pi(\mathbf r)\,\,d\mathbf r,
\end{equation}
where \(\hat H_{el}\) is defined in \Cref{eq:elec_ham}.
This identifies the \emph{local energy}
\begin{equation}
E_L(\mathbf r) = \frac{\hat H_{el}\Psi(\mathbf r)}{\Psi(\mathbf r)},
\end{equation}
which we can estimate by drawing \(\mathbf r^{(k)} \sim \pi\), yielding the Monte Carlo estimator of \(E[\Psi]\),
\begin{equation}
\widehat{E} = \frac{1}{M}\sum_{k=1}^{M} E_L\!\big(\mathbf r^{(k)}\big),
\label{eq:vmc_stoch_estimator}
\end{equation}
with variance reduced by efficient sampling. Optimizing the parameters of the trial wavefunction \(\Psi\) reduces this estimate to the ground-state energy \(E_0\) using the variational principle. Example trial wavefunctions include Jastrow-Slater \cite{talman1980jastrow, foulkes2001quantum}, multideterminant, backflow \cite{toulouse2007optimization} or neural network \pluralansatz as in Orbformer among others \cite{Hermann_2020,PhysRevX.14.021030, foster2025abinitiofoundationmodel}.  

We note that, unlike previous methods, VMC targets the many-electron wavefunction directly, thereby avoiding the choice of an exchange-correlation functional (DFT) and the single-determinant bias of CCSD(T). With a sufficiently expressive \singleansatz, especially neural-network wavefunctions optimized by VMC, we can capture both dynamic and static correlation, routinely matching or exceeding CCSD(T)\cite{Zhao2016VMCBetterThanCC1,Stella2011VMCBetterThanCC2,pfau2020ferminet,pfau2024ferminetexcitedstates}.
Moreover, the dominant costs scale roughly polynomially ($N^3-N^4$ in number of electrons) with extremely parallelisable sampling over the number of so-called walkers (i.e., electron samples). This leads to a drop in wall-time that scales almost linearly with compute.

The reliability and efficiency of VMC come from the choice of a sampling scheme that yields the stochastic estimator of \Cref{eq:vmc_stoch_estimator} (e.g., Markov-Chain Monte Carlo). These provide statistical error bars and systematic-bias controls, yielding calibrated predictions rather than point estimates.

\section{Large Wavefunction Models}
Much of VMC literature, specifically in quantum chemistry applications in the last few decades, has seen many changes and improvements for sampling techniques and stochastic estimators, but less in terms of the wavefunction \singleansatz itself. 
However, due to the rise in popularity and interest of ML in the quantum physics and quantum computing communities in recent years, a new class of expressive wavefunctions has arisen. We call these  Large Wavefunction Models (LWMs).

We can think of LWMs as the quantum analog of LLMs. In LLMs, a model with hundreds of millions to billions of parameters is \textit{pretrained} on vast amounts of text with a generic objective (e.g., next-token prediction), then \textit{fine-tuned} for a specific task, and finally run in \textit{inference} to answer prompts. Likewise, in LWMs, we build very large \pluralansatz that increase in expressivity as the number of model parameters grows (we have tested on the order of hundreds of millions of parameters across multiple GPUs) and \textit{pretrain} them on extensive datasets of molecular fragments that cover most chemical environments. Then, we \textit{fine-tune} to the target use case, such as multiple larger or more specific molecules and geometries.
At \textit{inference} (also named evaluation in this context), instead of generating tokens, the model is "queried" by specifying nuclear geometry, charges, boundary conditions, and electronic state. Analogously to an LLM, it produces an answer, except that here the answer is a quantitatively accurate, uncertainty-calibrated physical prediction of observables such as energies, forces, densities, and spectra. This concept is key to understanding where LWMs shine in terms of overall amortization of costs. Indeed, the typical use case of CCSD(T) or vanilla VMC calculations can be thought of as a combined process of \textit{pretraining}, \textit{fine-tuning} and \textit{inference} being done repeatedly for every single query system, with all the data computed being thrown away at every query and no ``knowledge" being shared across each computation. Training LWMs is underpinned by the same VMC training routine as NN-WFNs, across  various systems. In this sense, they are a generalization of NN-WFNs, which approximate the many-electron ground (and excited) states of a single system by representing the antisymmetric wavefunction $\Psi(\mathbf{R})$ with a flexible neural \singleansatz whose parameters $\theta$ are optimized by VMC. Due to the variational principle, the NN-WFN \singleansatz, bias and variance systematically improve and converge to the exact answer, producing highly accurate results. We will now briefly review the landscape of NN-WFNs.

Early fermionic NN-WFNs, such as FermiNet \cite{pfau2020ferminet}, established the core ideas. FermiNet builds generalized Slater determinants from learnable, permutation-equivariant electron features so that antisymmetry is exact by construction and correlation is captured through deep, non-local descriptors. Trained with VMC, it matched and often surpassed the CC accuracy on challenging small molecules and bond dissociations, demonstrating that deep networks can directly solve \textit{ab-initio} chemistry to high fidelity without an exchange-correlation functional. 
PauliNet took a complementary route, starting from physics-informed Jastrow-Slater structures and backflow-like coordinate transforms, enforcing electronic cusps and other exact constraints inside the network \cite{hermann2020paulinet}. Together, these works showed that neural \pluralansatz can encode fermionic antisymmetry, short-range electron–electron-/nuclear cusps, and mid-/long-range correlation within a single, trainable object. 

Subsequent advances broadened the scope and the technology. Neural backflow generalized the classic backflow transform with deep neural networks to improve nodal surfaces \cite{luo2019backneuralbackflow}. Self-attention/Transformer-based wavefunctions, such as Psiformer \cite{von2022psiformer}, replaced or augmented multi-layer perceptrons (MLPs) to capture long-range, many-body dependencies. NN-WFNs were also recently extended beyond finite molecules to periodic solids by embedding periodic boundary conditions and electron-ion equivariances directly into the \singleansatz \cite{wilson2023periodicnnwfn}.

A key limitation of the early NN-WFNs was the per-system training: each molecule required thousands of GPU-hours to optimize the variational parameters $\theta$ of the wavefunction. However, that barrier has recently been lifted by transferable and amortized approaches. Pioneering studies introduced parameter sharing and graph-learned orbital embeddings to generalize across related molecules, effectively training a wavefunction that is shared across different geometries (also called conformations) and chemical species \cite{gao2021pesnet1,gao2023generalizing,gao2024neuralpfaffian,scherbela2022weightsharing, scherbela2023finetuningnnwfn, scherbela2024taowfn, gerard2024transferablewfnsolids}. 

The most promising of these advances that made a larger-scale step is \microsoft's Orbformer, a chemically transferable NN-WFN (or, as we call it, LWM) that we make use of and improve in this work \cite{foster2025abinitiofoundationmodel}. Architecturally, Orbformer couples an \textit{Electron Transformer} (attention over electrons with nucleus-conditioned features) to an \textit{Orbital Generator} that produces localized, reusable generalized orbitals that promote locality, size extensivity and reuse across different chemical environments. 
The Orbformer model was trained variationally (i.e., without training labels) on $\sim$22k small chemical fragments and fine-tuned to new and larger systems, achieving sub $\sim$1 kcal/mol (i.e., ``chemical accuracy") on dissociation curves and Diels-Alder transition states while amortizing cost over multiple molecules and conformers. This makes Orbformer a foundation wavefunction that narrows the gap between bespoke VMC optimization and true pretrained inference. 
Although the Orbformer \singleansatz itself is expressive, the two main factors driving its cost are training and the choice of VMC sampling scheme. 

Indeed, in the following sections we show how the currently employed sampling schemes have limitations, in terms of quality and diversity of the electron samples, in making LWM future-proof (i.e. scalable to larger systems) and how more sophisticated sampling schemes, such as our proposed RELAX, can effectively reduce the cost of pretraining, fine-tuning, and evaluation procedures. In the next section, an introduction to stochastic estimators and a description of the technical improvements we contributed through this study are provided.

\section{Improvement of VMC Sampling Schemes}
In this section, we detail the steps employed in our novel scheme, Replica Exchange with Langevin Adaptive eXploration (RELAX). These steps significantly reduce the computational resources required to achieve chemical precision, enabling us to surpass Orbformer in effective sample size, autocorrelation, and amortization costs. We will begin by outlining two common sampling schemes and how we address their critical limitations, then proceed to describe RELAX.

\subsection{Limitations of Metropolis-Hastings and Langevin Sampling}

Recall from Section~\ref{sec:VMC} that VMC produces estimates for observables by sampling electronic configurations $\mathbf{r} = (\mathbf{r}_1, \ldots, \mathbf{r}_{N_c}$) drawn from the sampling distribution
\(\pi(\mathbf r)\propto |\Psi(\mathbf r)|^2\). To draw $\mathbf r \sim p(\mathbf r)\propto |\Psi(\mathbf r)|^2$, a common strategy is to employ Metropolis-Hastings (MH) and Langevin-based sampling algorithms. In MH one must choose a proposal kernel $q(\mathbf r'\!\mid\!\mathbf r)$ and accept the move $\mathbf r\!\to\!\mathbf r'$ with probability
\begin{equation}
\alpha(\mathbf r,\mathbf r')=\min\!\left(1,\frac{p(\mathbf r')\,q(\mathbf r\mid \mathbf r')}{p(\mathbf r)\,q(\mathbf r'\mid \mathbf r)}\right)
=\min\!\left(1,\frac{|\Psi(\mathbf r')|^2}{|\Psi(\mathbf r)|^2}\cdot\frac{q(\mathbf r\mid \mathbf r')}{q(\mathbf r'\mid \mathbf r)}\right).
\label{eq:mh-accept}
\end{equation}
Random-walk choices for $q$ (e.g., symmetric Gaussians) are readily implementable, but in high dimensions and near nodes/cusps they force a bad trade-off: large steps give low $\alpha$ and many rejections, whilst tiny steps give high $\alpha$ but microscopic movement. In either case, autocorrelation grows and the effective sample size shrinks.

A natural improvement is to switch to Langevin sampling. Using \(\pi(\mathbf r)\propto |\Psi(\mathbf r)|^2\), the overdamped Langevin SDE reads
\begin{equation}
d\mathbf r_t \;=\; \nabla_{\mathbf r}\log \pi(\mathbf r_t)\,dt \;+\; \sqrt{2}\,d\mathbf W_t
\;=\; \nabla_{\mathbf r}\log |\Psi(\mathbf r_t)|^2\,dt \;+\; \sqrt{2}\,d\mathbf W_t,
\label{eq:langevin-sde}
\end{equation}
whose stationary density is \(\pi\).
A single Euler--Maruyama step with stepsize \(\epsilon>0\) gives the unadjusted Langevin proposal (ULA)
\begin{equation}
\mathbf r' \;=\; \mathbf r \;+\; \epsilon\,\nabla\log |\Psi(\mathbf r)|^2 \;+\; \sqrt{2\epsilon}\,\boldsymbol\eta,
\qquad \boldsymbol\eta\sim\mathcal N(\mathbf 0,I).
\label{eq:ula}
\end{equation}
Because \Cref{eq:ula} discretizes \Cref{eq:langevin-sde}, it incurs a stepsize bias at finite \(\epsilon\).
To remove this bias while retaining the drift, we apply an MH correction with the asymmetric Gaussian proposal density,
\[
q(\mathbf r'\mid \mathbf r) \;=\; \mathcal N\!\big(\mathbf r'\,;\,\mathbf r+\epsilon\nabla\log|\Psi(\mathbf r)|^2,\;2\epsilon I\big),
\]
substituted into \Cref{eq:mh-accept}, yielding the Metropolis-Adjusted Langevin Algorithm (MALA) \cite{10.1063/1.436415, foster2025abinitiofoundationmodel}. This removes the bias while keeping the mixing gains of the Langevin proposals. 

However, MALA is also limited by two key factors. First, it incurs a high computational cost per accepted transition. This is because constructing the proposal and acceptance ratio requires evaluations of $|\Psi|$ and $\nabla \log |\Psi|^2$ at both the current and proposed states, so rejected moves waste expensive gradient computations. To mitigate rejection, one typically reduces the step size, at the cost of slowing exploration and increasing the integrated autocorrelation time, once again forcing us into a bad trade-off. Second, in electronic structure applications, the target exhibits severe anisotropy and stiffness (cusps, nodal surfaces), for which single-step Langevin dynamics with a scalar step size is poorly conditioned. Without an appropriate preconditioner or metric, MALA can revert to near–random-walk behavior in these regions, substantially degrading the effective sample size per second despite asymptotic exactness.


Our key innovation, a Replica Exchange with Langevin Adaptive eXploration (RELAX) algorithm, closes the gap posed by these methods. RELAX addresses these limitations by combining local and global transition kernels, which in turn significantly reduces the autocorrelation times compared to standard MALA. Our algorithm incorporates replica exchange \cite{URANO2015128,doi:10.1021/jp072655t} for better mixing and Riemannian metrics for geometry-aware proposals \cite{10.1111/j.1467-9868.2010.00765.x, NIPS2013_309928d4}. We will now describe these changes in more detail.

\subsection{Riemannian Langevin Dynamics}
Our Langevin proposal incorporates a Riemannian metric tensor $G(\mathbf{r})$, which preconditions the dynamics to take into account the local geometry of the configuration space. The update rule is an Euler discretization of the Riemannian Langevin SDE \cite{10.1111/j.1467-9868.2010.00765.x, NIPS2013_309928d4}

\begin{equation}
\mathrm{d}\circ r_{n+1}^i = \frac{1}{2} \left( G^{-1}(\mathbf{r}_n) \nabla_{\mathbf{r}} \log |\Psi(\mathbf{r}_n)|^2 \right)_i\mathrm{d}t +
\left( \sqrt{G^{-1}(\mathbf{r}_n)} \, \mathrm{d}\circ \mathrm{W}(t)\right)_i ,
\end{equation}
which we discretize, while accounting for the Itô-Stratonovich correction terms \cite{NIPS2013_309928d4} to ensure the invariant measure:
\begin{equation}
r_{n+1}^i = r_n^i + \epsilon \left( G^{-1}(\mathbf{r}_n) \nabla_{\mathbf{r}} \log |\Psi(\mathbf{r}_n)|^2 + \Gamma \right)_i
+  \left( \sqrt{2\epsilon G^{-1}(\mathbf{r}_n)} \, \boldsymbol{\eta}_n \right)_i,
\end{equation}
where,
\[
\Gamma_i = \sum_{j=1}^D \left( G^{-1}(\mathbf{r}_n) \right)_{ij} \operatorname{Tr} \left( G^{-1}(\mathbf{r}_n) \frac{\partial G(\mathbf{r}_n)}{\partial r^j} \right) - \sum_{j=1}^D \left( G^{-1}(\mathbf{r}_n) \frac{\partial G(\mathbf{r}_n)}{\partial r^j} G^{-1}(\mathbf{r}_n) \right)_{ij}.
\]
Here, $\boldsymbol{\eta}_n \sim \mathcal{N}(\mathbf{0}, I_D)$, $\epsilon$ is the step size adapted to ensure the desired acceptance rate, and $d$ is the dimension of the configuration space ($d= 3n_{electrons}$). The terms involving $\partial G / \partial r^j$ are the Itô-Stratonovich correction terms that account for the curvature of the manifold and preserve the target invariant distribution $|\Psi(\mathbf{r})|^2 \, d\mathbf{r}$. 

We note that it is not strictly necessary to implement the Ito-Stratonovich term in the presence of rejection adjustment.
However, we observed a consistent degradation of performance without it because electrons that are near nuclei fall onto nuclei without such a correction.

We chose a position- and species-dependent metric tensor $G$ that depends on the positions of the electrons, nuclei, the atomic numbers, and the spins to capture the curvature of the wavefunction. This yields proposals that are larger and farther away from the nuclei\footnote{ Implementation specifics are beyond the scope of this report; processed code and analysis are available from the corresponding authors on reasonable request (subject to standard agreements)}. In our benchmarks, this reduced the integrated autocorrelation time by up to 2x with negligible overhead.


This is followed by a Metropolis-Hastings adjustment to ensure detailed balance:
\begin{equation}
\alpha = \min\left(1, \exp\left( \log |\Psi(\mathbf{r}')|^2 - \log |\Psi(\mathbf{r})|^2 + \log q(\mathbf{r} | \mathbf{r}') - \log q(\mathbf{r}' | \mathbf{r}) \right) \right),
\end{equation}
where $q(\cdot | \cdot)$ is the density of the proposal that accounts for the metric.

\subsection{Replica Exchange}
To enhance mixing, we employ deterministic replica exchange between $K$ replicas at inverse temperatures $\beta_1 > \beta_2 > \cdots > \beta_K = 1$ (where $\beta=1$ corresponds to the target distribution) so that each replica $i$ targets $\pi_{\beta_{i}} = \pi_{0}^{1-\beta_{i}}(\mathbf{r}_{i})|\Psi(\mathbf{r}_{i})|^{2\beta_{i}}$. For every fixed number of steps, we try to swap between adjacent replicas $i$ and $i+1$ with acceptance probability \cite{URANO2015128,doi:10.1021/jp072655t}:
\begin{equation}
\begin{aligned}
\alpha_{\text{swap}}
= \min\Bigl(1,\; \exp\bigl( (\beta_{i+1}-\beta_{i}) (&\log \pi_{0}(\mathbf r_{i+1})
-\log \pi_{0}(\mathbf r_{i}) \\
&\qquad\quad + \log |\Psi(\mathbf r_{i})|^{2}
-\log |\Psi(\mathbf r_{i+1})|^{2}) \bigr) \Bigr).
\end{aligned}
\end{equation}
where $\mathbf{r}_i$ is the configuration of replica $i$. The deterministic aspect refers to a fixed swapping schedule, ensuring reproducibility and optimized exploration that minimizes the path length required for a sample to move from a zero temperature replica to a target replica.

In addition, temperatures are adjusted online to ensure a uniform acceptance probability among neighboring replicas, which in turn leads to minimization of the path lengths of swaps, resulting in theoretically optimal mixing \footnote{Details of the implementation of online adjustment are beyond the scope of this pre-print; researchers interested in reproducing these results may contact the corresponding author—processed code and analysis are available on reasonable request (subject to appropriate agreements).}.

\subsection{Global Replacement Kernel}
Additionally, our MCMC includes a global step \cite{samsonov2022localglobalmcmckernelsbest} that periodically proposes new configurations from a reference distribution $q(\mathbf{r})$, which is the same reference used in the replica of zero-temperature ($\beta = 0$) for efficient cold-chain sampling. This $q(\mathbf{r})$ is a mixture of exponential distributions centered on nuclear positions $\mathbf{R}_A$ and accounts for the structures of the atomic shell,
\begin{equation}
\pi_{0}(\mathbf{r}) = \sum_{A=1}^{N_{\text{atoms}}} \sum_{s=1}^{S_A} w_{A,s} \, \exp\left( -\alpha_{A,s} \, |\mathbf{r} - \mathbf{R}_A| \right),
\end{equation}
where $w_{A,s}$ are weights, $\alpha_{A,s}$ are decay parameters for atom $A$ and shell $s$, and $S_A$ is the number of shells per atom. This distribution can be sampled analytically and efficiently. However, our choice of $w_{A, s}$, nor $\alpha_{A,s}$, and the associated efficient sampling scheme are not detailed in this version; processed code and pipelines are available from the corresponding author on reasonable request (subject to standard agreements).

We note that all known research applications of such a kernel in the VMC setting use Gaussian mixtures without accounting for the electron shell structure. In our experiments, we observed that the choice of a Gaussian mixture leads to systematic undersampling of tails and, thus, to systematically overestimated energies.

The global proposal replaces the current configuration with a new sample from $q$, accepted with:
\begin{equation}
\alpha_{\text{global}} = \min\left(1, \frac{|\Psi(\mathbf{r}')|^2 / q(\mathbf{r}')}{|\Psi(\mathbf{r})|^2 / q(\mathbf{r})} \right),
\end{equation}
acting as an independence sampler to prevent trapping in local modes. The zero-temperature replica has acceptance 1 because its proposal matches its target. With an optimal temperature schedule, the swap moves create regeneration, replacing the state of each chain in a finite expected time.


\section{Results \& Discussion}
In this section, we begin by describing the setup used to obtain the benchmarks, then detail our method for cost analysis of VMC data generation. We then discuss the autocorrelation time, effective sample size, and cost reduction in \simulacra's pipeline, contrasting it with \microsoft's framework \cite{foster2025abinitiofoundationmodel}.
\subsection{Experimental setup}
All experiments in this study were performed with our proprietary pipeline for the implementation of RELAX and with the open-source \textit{oneqmc} framework for experiments labelled \microsoft \cite{oneqmc2025} throughout the analysis. The latter's experiments were run with default settings.
The amino acids geometries (in their L-chiral form) were obtained through conformer generation (ETKDGv3) using RDKit \cite{rdkit}. Hydrogen atoms were added to have all systems with a total charge of zero (neutral) and a spin multiplicity of 1. Note that because Orbformer's pretrained model was trained on chemical species up to the fluorine atom (Z=9), only a few natural amino acids were selected for this study. The selection was done based on the increasing number of atoms and electrons, up to TRP (tryptophan) with 108 electrons and 27 atoms.
Note that we include the methane ($\mathrm{CH_4}$) molecule as a baseline of a small number of particles (=10) and atoms (=5).

All experiments were run on one single GPU (H200) for comparison; thus, no scaling in terms of number of GPUs is portrayed in this work. Calculations for the CC time estimates were done using PySCF(v2.1)\cite{pyscf} by computing energies at the CCSD(T)/CBS level of theory using Restricted HF wavefunctions, with the 2-point CBS extrapolation performed through cc-pVTZ and cc-pVQZ basis-sets calculations. PySCF calculations were performed on 8 CPU cores and 128GB of RAM. 
\subsection{Cost Analysis for VMC Data Generation}
Monte-Carlo sampling based on Markov Chains (MCMC) has two key factors that quantify the number of samples needed to reach an error level $\epsilon = 10^{-3}$ for energy estimation. The goal is to obtain $\propto\frac{1}{\epsilon^2}$ statistically independent samples. However, in MCMC sampling, successive samples are often correlated, which means they do not provide independent statistical information. We can measure the statistical dependence between successive samples in the Markov chain with autocorrelation; lower autocorrelation values indicate more independent samples. The autocorrelation time $\tau := \sum_{t\ge 1} \mathrm{corr}(\mathbf{r}^{(0)}, \mathbf{r}^{(t)})$ \cite{bialas2023analysis} then quantifies how many sampling steps are needed, on average, to generate an effectively independent sample. 
Consequently, the total number of samples to reach an error $\epsilon$ increases by a factor $\propto \frac{\tau}{\epsilon^2}$ samples to achieve the same error \cite{10.1093/oso/9780198522669.003.0038} as an unbiased estimator that is autocorrelation-free (which is never the case in MCMC schemes). Given the lack of access to infinite series, we opt for the AR(1) approximation of autocorrelation time: $\tau \approx \frac{1+ \mathrm{corr}(\mathbf{r}^{(0)}, \mathbf{r}^{(1)})}{1- \mathrm{corr}(\mathbf{r}^{(0)}, \mathbf{r}^{(1)})}$, which is reasonable in our context, given that we are dealing with Markov chains and thus autocorrelation is strongly dependent on the spectral gap (i.e, the inverse of the spectral gap is proportional, up to constants, to the integrated autocorrelation time and to the mixing time) of the Markov operator, making the AR(1) approximation asymptotically justified.

With that in mind, the cost to generate a dataset of energies through an LWM and VMC per single data set point can be decomposed as:
\begin{equation}
C = \frac{C_{\text{pretrain}}}{D} + \frac{C_{\text{fine-tune}}}{K} + C_{\text{eval}},
\end{equation}
where:
\begin{itemize}
\item $C_{\text{pretrain}}$ is \textbf{amortized} over the size of the dataset ($D$), which tends to be of the order of at least $10^{4}-10^{5}$ data points (i.e. different molecules and geometries),
\item $C_{\text{fine-tune}}$ is \textbf{amortized} over the number of conformations per molecule ($K$) that tends to be on the order of $10^{2}-10^{3}$ data points,
\item $C_{\text{eval}}$ \textbf{dominates for large datasets and is the focus here.}
\end{itemize}
All three terms $C_{\text{pretrain}}$, $C_{\text{fine-tune}}$, $C_{\text{eval}}$ scale proportionally with autocorrelation time $\tau$ and forward pass time $t_f$. Thus, we treat pretraining as an upfront one-off computational investment and focus the discussion on inference performance. Moreover, we expect finetuning to diminish with scaling laws as pretraining larger models will allow us to achieve near-zero-shot evaluation (i.e., inference) performance due to the growing expressive nature of the underlying wavefunction \singleansatz as number of parameters grows \cite{Augustine2024UAT}. 

For evaluation, the cost to achieve energy error $\epsilon = 10^{-3}$ Ha is:
\begin{equation}
C_{\text{eval}} \propto \sigma_{0} \cdot (t_E + t_{\text{MCMC}}) \cdot \tau / \epsilon^2,
\end{equation}
where $\sigma_{0}$ is the standard deviation of energy, $t_E$ is the energy computation time (scaling as $\mathcal{O}(N^4)$ for electron-electron terms in some \pluralansatz), $t_{\text{MCMC}}$ is the MCMC proposal time ($\mathcal{O}(N^3)$), $\tau$ is autocorrelation time, and $t_f$ is forward pass time. The $1/\epsilon^2$ arises from the need for independent samples $N_{\text{indep}} \approx (\sigma / \epsilon)^2$, where $\sigma$ is the standard deviation of $E_L$ (local-energy) and the total steps $\approx \tau \cdot N_{\text{indep}}$.

\subsection{Autocorrelation Time}



\begin{figure}[hbt!]
    \centering
    \includegraphics[width=\textwidth]{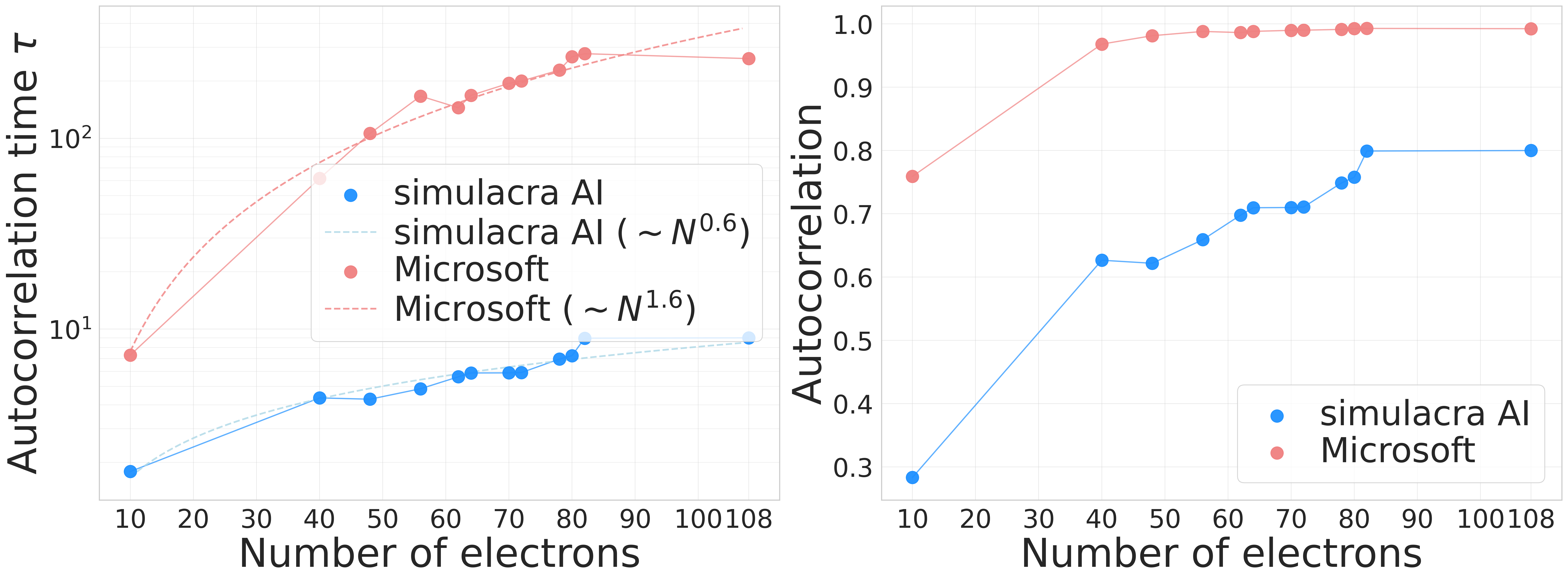}
    \caption{\textbf{Autocorrelation time $\tau$ and value} for different systems. Lower values are better. \simulacra maintains $\tau < 10$ and autocorrelation $\sim$0.7-0.8, producing largely independent samples, while \microsoft exhibits poor scaling with $\tau > 100$ and autocorrelation $\sim$0.95-1.0, indicating highly correlated samples that provide minimal additional statistical information.}
    \label{fig:autocorr_time}
\end{figure}

We benchmark autocorrelation times and values across different systems using 1024 walkers and 10,000 evaluation steps per system.  
\Cref{fig:autocorr_time} demonstrates the substantial improvement achieved by RELAX sampling: \simulacra's pipeline maintains autocorrelation times of $\tau \approx 5-10$  and autocorrelation values of 0.65-0.80 across all system sizes, indicating that samples become largely independent after relatively few iterations. In contrast, \microsoft's implementation exhibits higher autocorrelation times reaching $\tau > 150$ with autocorrelation values approaching unity ($\sim$0.95-1.0) for systems with $>$30 electrons, indicating highly correlated samples that provide minimal additional statistical information. This improvement stems from RELAX's efficient replica exchange mechanism using 8 steps per iteration across 8 replicas, compared to 60 MALA steps in \microsoft's approach. Generating more independent samples directly translates into higher effective sample sizes, enabling convergence with fewer total sampling steps and reducing overall computational cost, even when per-epoch times are sometimes longer.
Looking at \Cref{fig:autocorr_time} left we can also observe that $\tau$ grows sub-linearly with system size in our pipeline ($\mathcal{O}(N_{electrons}^{0.6})$), while it explodes in \microsoft's Orbformer pipeline at least as $\mathcal{O}(N_{electrons}^{1.6})$ based on power law fit. This implies that the total theoretical complexity of VMC for \simulacra pipeline is around $\mathcal{O}(N_{electrons}^{4.6})$, while for \microsoft, it is at least $\mathcal{O}(N_{electrons}^{5.6})$. Here, latter takes into account the theoretical lower-bound $\mathcal{O}(N_{electrons}^{4})$ scaling of VMC. This analysis is performed only on evaluation, so we expect the fine-tuning to be potentially even worse \cite{scherbela2025accurateabinitioneuralnetworksolutions}. 

\subsection{Effective Sample Size \& Cost Reduction}
\begin{figure}[hbt!]
    \centering
    \includegraphics[width=\textwidth]{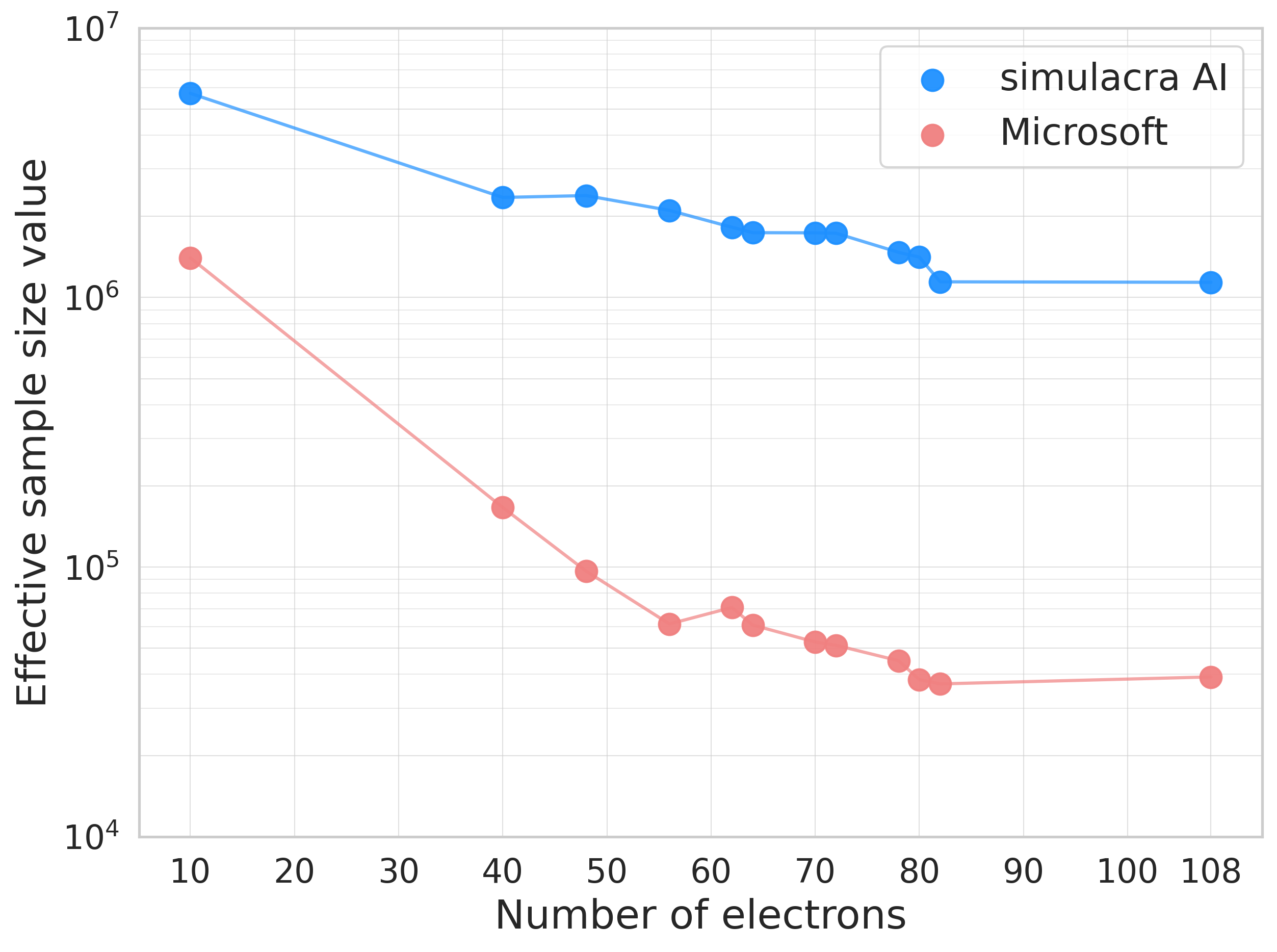} 
    \caption{\textbf{Effective sample size} for $10^4$ iterations with 1024 walkers. ESS quantifies the number of statistically independent samples produced; higher values indicate better sampling efficiency. \simulacra's RELAX (blue) maintains ESS of $\sim 1.2-6 \times 10^6$ across all system sizes, while standard MALA (red) has much less ($<10^5$) independent samples for large systems due to severe autocorrelation.}
    \label{fig:ess}
\end{figure}

We note that there is a systematic reluctance in the literature to account for autocorrelation time correction in analyses of VMC, likely because, up until now, most research has assumed the computation of one system at a time.
The effective sample size (ESS) quantifies the number of statistically independent samples produced by an MCMC sampler from a given number of iterations. Higher ESS indicates more efficient sampling, as each iteration contributes more unique statistical information. \Cref{fig:ess} compares the ESS achieved by both pipelines for a fixed computational budget of $10^{4}$ iterations with 1024 walkers. \simulacra's RELAX sampling produces ESS values ranging from $1.2 \times 10^6$ for the largest system (108 electrons) to $5.7 \times 10^6$ for the smallest system (10 electrons), demonstrating effective generation of independent samples even as system size increases. In contrast, \microsoft's MALA implementation shows dramatically lower ESS values, dropping from $1.4 \times 10^6$ to  $<10^5$ for systems with $>50$ electrons. This difference directly reflects the autocorrelation behavior shown in \cref{fig:autocorr_time}: \microsoft's high autocorrelation times result in highly correlated samples that contribute minimal additional statistical information, effectively wasting computational resources. The superior ESS of RELAX directly leads to faster convergence and reduced computational cost, as fewer total iterations are required to achieve the target statistical precision.

\begin{table}[ht]
\centering
\scalebox{0.8}{
\begin{tabular}{lccccccc}
\toprule
Molecule & $n_{\text{atoms}}$ & $t_{\text{\microsoft}}$ & $t_{\text{\simulacra}}$ & $\tau_{\text{\microsoft}}$ & $\tau_{\text{\simulacra}}$ & Ratio $R$  & \begin{tabular}{@{}c@{}} Cost \\ Reduction \% \end{tabular} \\
\midrule
Methane & 5 & 0.40 & 0.24 & 7.2981 & 1.7904 & 7.0269 & \textbf{-85.7}\\
Glycine & 10 & 2.11 & 2.17 & 61.5784 & 4.3558 & 46.1948 & \textbf{-97.8}\\
Alanine & 13 & 2.78 & 2.77 & 106.0242 & 4.2874 & 23.9650 & \textbf{-95.8}\\
Serine & 14 & 3.44 & 3.45 & 166.4675 & 4.8649 & 54.2185 & \textbf{-98.1}\\
Proline & 17 & 4.37 & 4.21 & 144.6248 & 5.6191 & 33.6757 & \textbf{-97.0}\\
Valine & 19 & 4.49 & 4.45 & 168.0845 & 5.8853 & 34.8227 & \textbf{-97.1}\\
Aspartic Acid & 16 & 35.74 & 48.43 & 194.2190 & 5.8929 & 23.0330 & \textbf{-95.6}\\
Leucine & 22 & 24.84 & 49.81 & 199.7660 & 5.9145 & 17.9397 & \textbf{-94.4}\\
Glutamine & 20 & 27.19 & 53.45 & 227.8487 & 6.9612 & 26.0761 & \textbf{-96.1}\\
Lysine & 24 & 28.18 & 54.47 & 267.7938 & 7.2528 & 22.9917 & \textbf{-95.6}\\
Histidine & 20 & 28.89 & 55.95 & 277.7712 & 8.9557 & 24.5316 & \textbf{-95.9}\\
Tryptophan & 27 & 43.80 & 73.18 & 261.7089 & 8.9941 & 22.3448 & \textbf{-95.5}\\
\bottomrule
\end{tabular}}
\caption{\textbf{Benchmark results} for 12 systems of increasing number of atoms and electrons. The time $t$ of a pipeline's epoch is reported in seconds ($s$) and averaged across the 10k evaluation steps of each run. The average ratio $R$ across all systems is $\approx 28$. The cost reduction is calculated as $=100\% * (1-1/R)$.}
\label{tab:results}
\end{table}

\subsection{Scaling Laws \& Cost Reduction}\label{sec:scaling_laws_results}
 To compare the cost reduction gain between our pipeline against \microsoft's reference pipeline and to show the overall gain of our optimizations and algorithmic advances (e.g., RELAX), we define the cost ratio (\simulacra / \microsoft) $R$ to achieve $\epsilon = 10^{-3}$ as
\begin{equation}
R = \frac{\tau_{\text{\microsoft}} \cdot t_{\text{\microsoft}} \cdot \sigma_{\text{\microsoft}}^2}{\tau_{\text{\simulacra}} \cdot t_{\text{\simulacra}} \cdot \sigma_{\text{\simulacra}}^2},
\end{equation}
which takes into account both statistical measures (autocorrelation time) and speed ($t$) of an iteration (in this case, an epoch). 
The results  are shown in \Cref{tab:results}. A mean Ratio $R$ of $\sim$28 is achieved, showing a significant increase in efficiency in our pipeline across all system sizes.
On the other hand, to illustrate the efficiency gains of neural network-based VMC, we compare the cost scaling of both pipelines against traditional CCSD methods, which scale as $O(N_{basis}^6)$ with $N_{basis}$ the number of one-particle basis functions. Cost estimates were based on current hardware pricing: $\sim$\$2.20 per GPU-hour for NVIDIA H200 (used for VMC pipelines), \$0.007 per CPU-hour for PySCF implementations, and \$0.001 per CPU-hour for ORCA \cite{https://doi.org/10.1002/wcms.1606} based on their platform pricing. CCSD computational times were obtained from both ORCA single-point calculations and our PySCF benchmarks. To ensure conservative estimates, we deliberately underestimate CCSD costs and present them as a range that spans different implementations and hardware configurations.

For these pipelines, costs are calculated to achieve an energy precision of $10^{-3}$ Ha, assuming standard errors in the range $10^{-1}\le \sigma\le 10^{0}$ Ha. This conservative approach accounts for potential fine-tuning costs, though these are not included in the direct comparison. 

\begin{figure}[htb!]
    \centering
    \includegraphics[width=\textwidth]{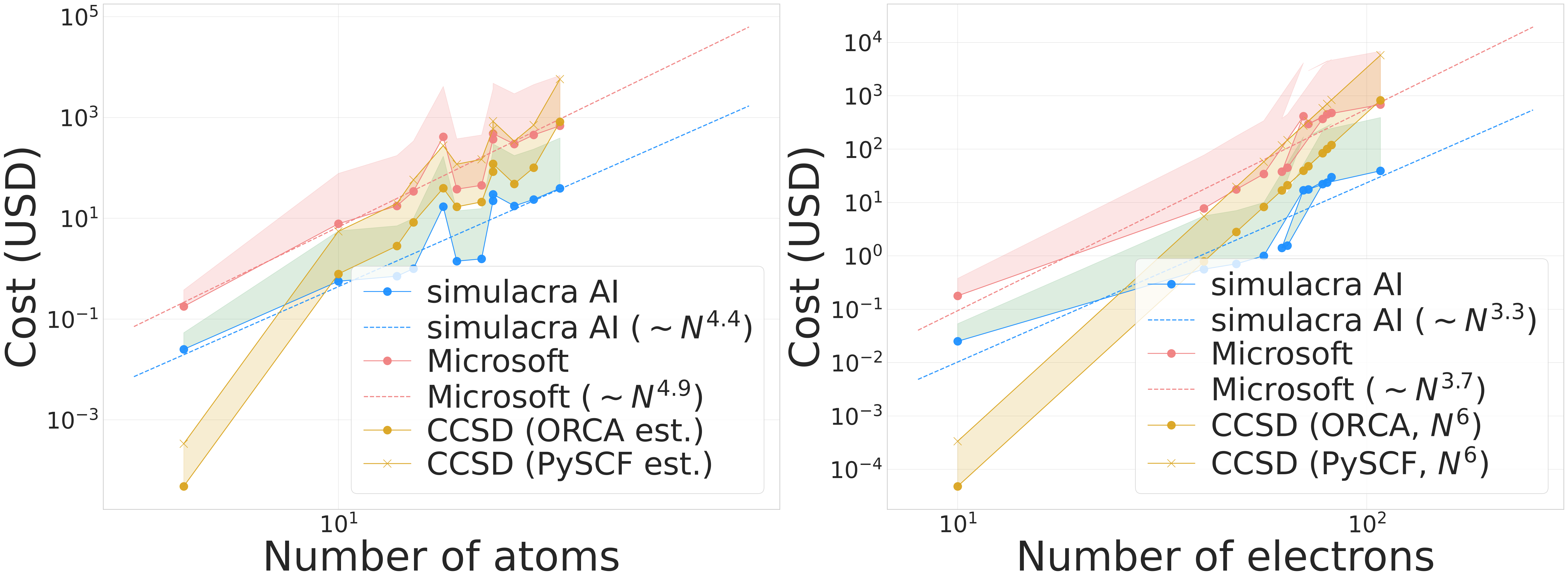}
    \caption{\textbf{Cost per system} (in USD) depending on the number of atoms and the number of electrons. \simulacra AI's pipeline achieves 15-50$\times$ cost reduction relative to \microsoft's one across all system sizes, with improved power-law scaling ($N_{atoms}^{4.4}$ vs. $N_{atoms}^{4.9}$; $N_{electrons}^{3.3}$ vs. $N_{electrons}^{3.7}$). Both VMC approaches significantly outperform CCSD methods for systems with $\ge 10$ atoms, with cost advantages increasing for larger molecules.}
    \label{fig:costs}
\end{figure}

As shown in \Cref{fig:costs}, \simulacra's framework achieves cost parity with CCSD for the Glycine system (10 atoms, 40 electrons). It also demonstrates 2-3$\times$ cost reduction relative to CCSD for any other system larger than 10 atoms.\footnote{Assuming large-scale dataset generation rather than single-system calculations.}The favorable power-law scaling of $N_{electrons}^{3.3}$ vs. CCSD's $N_{electrons}^6$ projects order-of-magnitude cost reductions for systems with $\ge 32$ atoms. Note that our estimates for CCSD assume that $N_{electrons} = N_{basis}$, which is an ideal lower-bound for the CCSD method, but in practice, the number of one-particle basis functions is generally much larger than the number of electrons by at least a few orders of magnitude. This realistically means even better practical efficiency for our framework. 
Moreover, \simulacra's pipeline maintains 15-50$\times$ cost advantages over \microsoft's Orbformer implementation across all system sizes, with the gap widening for larger molecules due to superior scaling exponents $N_{atoms}^{4.4}$ vs. $N_{atoms}^{4.9}$ and $N_{electrons}^{3.3}$ vs. $N_{electrons}^{3.7}$, respectively.

\section{Conclusion}

We have shown how \simulacra's VMC pipeline leverages this scheme to reduce data generation costs by $\sim$15-50x compared to \microsoft's Orbformer pipeline (on average 28x) and $\sim$2-3x compared to traditional CC methods for any molecule larger than 10 atoms. Moreover, we have observed that our RELAX autocorrelation times for evaluation (i.e., inference) scales as $\mathcal{O}(N_{electrons}^{0.6})$ compared to the $\mathcal{O}(N_{electrons}^{1.6})$ MALA implementation in \microsoft's pipeline. This clearly shows that RELAX is favorable for scaling to larger and larger systems.
Compared to traditional CC methods, which scale as $\mathcal{O}(N_{basis}^{6})$ to $\mathcal{O}(N_{basis}^{7})$, we have shown that the Orbformer LWM paired with proprietary RELAX sampling scales in practice as $\sim N_{electrons}^{3.3}$ while maintaining or even outperforming the accuracy of CC. This is remarkably better than the theoretical lower bound of $N^{4+\omega}$ (where $\omega$ comes from MCMC $\tau$ scaling). The improvement in scalability is due to the high parallelization performed on GPU accelerators.

There is a common misconception in the VMC community that MCMC algorithms can sidestep the curse of dimensionality in many-body systems. This is because people assume that such sampling procedures sidestep diagonalization with sampling-based estimators. However, this neglects the fact that, to obtain statistically independent samples and ensure unbiased estimators, we need to run the MCMC scheme past its autocorrelation time. The autocorrelation time for existing schemes in the literature scales with the spectral gap of the proposal densities, and such a spectral gap scales \textit{exponentially} in the system size (which is already exponential at the level of Hilbert spaces) for these existing schemes \cite{andrieu2015convergence}. This means that the curse of dimensionality is not removed, it is recast in the sampling scheme with no real gains on the scales that matter for the pharmaceutical sector \cite{raginsky2017non}. However, our sampling scheme \textit{genuinely bypasses the curse}, as it decouples the size of the spectral gap and the system dimension \cite{dong2022spectral}. We therefore anticipate that the gains shown on these test cases only to improve as we scale to larger systems. This paves the way for the creation of affordable, extremely accurate, and large-scale ab initio datasets, accelerating AI-driven discoveries in the pharmaceutical industry and beyond.  

\section{Code Availability}
The pretrained model used in this study can be found at \url{https://github.com/microsoft/oneqmc}. The pipeline developed for this work is not publicly available, but collaborative enquiries are welcome. Interested researchers may contact the corresponding authors.
\section{Acknowledgments}
This project was supported by the resources and services provided by Nebius, the Google Cloud Startup Program, and the NVIDIA Inception Program.
\section{Declarations}
This study refers to \microsoft only to identify previously published methods used as comparators. The authors have no affiliation with \microsoft, and no funding, materials, or assistance were received from the company. Reference to third-party names does not imply endorsement.
Aleksei Ustimenko, Elena Orlova, Philip Mantrov, Fabio Falcioni, and Timothy Heightman are shareholders of Simulacra Research Inc.

\printbibliography
\end{document}

%% file: utils.tex
\definecolor{orange}{HTML}{F28C28}

\usepackage{xspace}
\usepackage{roboto} 
\newcommand{\pluralansatz}{\textit{ansätze}\xspace}
\newcommand{\singleansatz}{\textit{ansatz}\xspace}
\newcommand{\simulacra}{{\fontfamily{Roboto-TLF}\selectfont simulacra AI}\xspace}\newcommand{\microsoft}{\textit{Microsoft}\xspace}